\documentclass[aps, prc, floatfix, nofootinbib, superscriptaddress, twocolumn]{revtex4-1}

\usepackage{latexsym}
\usepackage{amsmath}
\usepackage{amssymb}
\usepackage{amsfonts}

\usepackage[mathscr,scaled=1.15]{urwchancal}
\DeclareFontFamily{OT1}{pzc}{}
\DeclareFontShape{OT1}{pzc}{m}{it}%
{<-> s * [1.15] pzcmi7t}{}
\DeclareMathAlphabet{\mathpzc}{OT1}{pzc}{m}{it}

\usepackage{color}

\usepackage{supertabular}
\usepackage{placeins}
\usepackage{epsfig}
\usepackage{graphicx}
\usepackage{booktabs}

\definecolor{purple}{rgb}{0.5,0,0.5}
\definecolor{blue}{rgb}{0.0,0,0.9}
\definecolor{prdblue}{rgb}{0.133,0.118,0.498}
\usepackage[colorlinks=true, pdfstartview=FitV, linkcolor=prdblue, citecolor= prdblue, urlcolor=prdblue]{hyperref}

\begin{document}

\title{The light quarkonium and charmonium mass shifts in an unquenched quark model}

\author{Xiaoyun Chen}
\email[]{xychen@jit.edu.cn}
\affiliation{College of Science, Jinling Institute of Technology, Nanjing 211169, P. R. China}

\author{Yue Tan}
\email[]{tanyue@ycit.edu.cn}
\affiliation{Department of Physics, Yancheng Institute of Technology, Yancheng 224000, P. R. China}

\begin{abstract}
The unquenched quark model for the light quarkonium and the charmonium states is explored in the present work. The quark-pair creation operator in the $^3P_0$ model,
which mix the two-quark and four-quark components is modified by considering the effects of the created quark pair's energy, as well as the separation between the
created quark pair and the valence quark pair. All the wave functions needed including the mesons and the relative motion between two mesons are all obtained by
solving the corresponding Schr\"{o}dinger equation with the help of the Gaussian expansion method. Our aim of the present work is to find a new set of parameters
which can give a good description of the mass spectrum of the low-lying light quarkonium and charmonium states. Moreover, some exotic states, for example $X(3872)$
can be described well in the unquenched quark model.

\end{abstract}


\maketitle


\section{Introduction} \label{introduction}
In a nonrelativistic valence quark model, a baryon is consisted of three quarks and a meson is composed of quark-antiquark. The model has been described the properties of the low-lying hadrons and hadron-hadron interactions successfully. For example it is successfully applied to heavy quarkonia, such as bottomonium and charmonium \cite{Eichten:1978tg, Eichten:1979ms, Gupta:1982kp, Gupta:1983we, Gupta:1984jb, Gupta:1984um, Kwong:1987ak, Kwong:1988ae, Barnes:1996ff, Ebert:2002pp, Radford:2007vd, Eichten:2007qx, Segovia:2008zz, Danilkin:2009hr, Ferretti:2013vua, Segovia:2013wma, Godfrey:2015dia, Segovia:2016xqb}, and also, to some extent, light mesons \cite{Godfrey:1985xj, Vijande:2004he, Segovia:2008zza}.
With the progresses of experiments, more and more new exotic hadrons have been reported by experimental collaborations since 2003,
these exotic states cannot be described well by the valence quark model, which poses a great challenge for the quark model.

For example, the measured mass of the second $P$-wave charmonium state $X(3872)$ \cite{Belle:2003nnu} is 100 MeV lower than the predicted mass by the quark model for $\chi_{c_1}(2P)$,
and its decay width is $\le$ 1 MeV. Similar problems are also observed for the charmed meson states $D_{s_0}^*(2317)$ \cite{BaBar:2003oey} and $D_{s_1} (2460)$ \cite{CLEO:2003ggt}.
These puzzling issues have led theorists to refer to them as "exotic states". Various explanations, such as multi-quark states, hybrid states, gluonic excitations, etc. are proposed.

Actually, in order to describe these exotic hadrons in the quark model, the model has to be extended. By considering that the quark number is not a conserved quantity and the quark pairs
$q\bar{q}$ can be excited in the vacuum, a new quark model, the unquenched quark model (UQM) so called is developed. The wave functions of meson and baryon in UQM can be written as follows,
\begin{align}
&|\mbox{Meson}>=|q\bar{q}\rangle+|q\bar{q}q\bar{q}\rangle+|q\bar{q}g\rangle+... \\
&|\mbox{Baryon}>=|qqq\rangle+|qqqq\bar{q}\rangle+|qqqg\rangle+...
\end{align}
The first item represents the wave function in the nonrelativistic valence quark model. The second and the third item consider the quark pairs and the gluon excitation in the vacuum.
As a preliminary phase of the development of UQM, only the first two terms, the valence term and the valence with quark-antiquark excitation are taken into account in the model.

So far, there have been many theoretical studies exploring the effects of quark-antiquark pair excitation on the properties of hadrons. For example, in a recent article published
in ``Nature" \cite{SeaQuest:2021zxb}, scientists presented the asymmetry in the momentum distribution of antimatter quarks, indicating evidence of matter-antimatter asymmetry within
the proton. Regarding the Roper resonance $N(1440)$, the latest results suggest that it is a radially excited state of the proton core surrounded by a $20\%$ meson cloud \cite{Burkert:2017djo}.
Furthermore, by considering meson-baryon coupling effects, Kenta Miyahara \emph{et al.} proposed that $\Lambda(1405)$ is a mixture of three-quark and five-quark states,
with $\bar{K}N$ being the dominant component \cite{Miyahara:2020jmn}.

For heavy-light systems, Beveren \emph{et al.} considered the $DK$ coupling channel effects in the $c\bar{s}$ system and performed calculations on the mass of $D_{s_0}^{*}(2317)$,
which provided a good explanation for the experimental data \cite{vanBeveren:2003kd}. In the work of Albaladejo \emph{et al.} \cite{Albaladejo:2018mhb}, the influence of $D^{(*)}K$
meson-meson coupling channels on $P$-wave $c\bar{s}$ states were taken into account to study the internal structure of $D_{s_0}^*(2317)$ and $D_{s_1}(2460)$. They suggested that these particles
are predominantly composed of a four-quark structure mixed with a quark-antiquark component.

For the heavy systems, by considering coupling channel effects, the mass of charmonium state $\chi_{c_1}(2P)$ can be lowered to the value of
$X(3872)$ \cite{Kalashnikova:2005ui,Zhang:2009bv,Li:2009zu,Braaten:2007ft,Ebert:2008kb,Ferretti:2014xqa,Tan:2019qwe}. In the study of $\psi(4415)$, Cao and Zhao took into account
the influence of molecular states $D_{s_1}\bar{D}_s$ and $D_{s_0}\bar{D}_s^*$ in the unitarized picture \cite{Cao:2017lui}. Luo \emph{et al.} calculated the mass spectrum of
$\Lambda_{c}(2P,(3/2)^-)$ by considering the coupling channel effects of $D^*N$, which provided a good explanation for the charmed baryon state $\Lambda_c(2940)^+$ reported
by BaBar Collaboration \cite{Luo:2019qkm}. Furthermore, the possibility of placing the $X(3915)$, which is produced through the two-photon fusion process, as a charmonium family member
$\chi_{c_0} (2P)$ is closely related to the coupling channel effects.

These researches prompt us to continue delving into and developing the unquenched quark model. Recently, this has been a crucial topic in the field of hadron physics,
accompanied by the discovery of numerous new hadronic states and the accumulation of relevant experimental data. Generally, the transition operator which mix the quark-antiquark and
four-quark components is taken from the $^3P_0$ model in these theoretical calculations. Some of the previous work found that the virtual quark pair creation in hadronic system leads
to a very large mass shifts \cite{Barnes:2007xu,Chen:2017mug}. The large mass shift will challenge the validity of the valence quark model in describing the ground state hadrons and
the convergence of UQM. The convergence problem was also noted by Ferretti and Santopinto, it is tackled by considering only the contribution from the closest set of meson-meson
intermediate states and taking the contribution from other states as some kind of global constant \cite{Ferretti:2018tco}. In our previous work \cite{Chen:2017mug,Chen:2023wfp},
we try to solve this problem by modifying the transition operator, i.e. introducing energy and separation damping factors. With the improved transition operator, the mass shifts of
the low-lying light mesons \cite{Chen:2017mug} and charmonium \cite{Chen:2023wfp} are significantly reduced.  The proportion of the two-quark component rises to around $90\%$,
suppressing the influence of the four-quark components. This ensures the validity of the constituent valence quark model in describing the low-lying hadron states.

Because of incorporating the four-quark components, the model parameters used in the valence quark model have to be adjusted. In the present work, with the improved transition operator,
the meson spectrum is computed by solving the eigenequation of the unquenched quark model Hamiltonian. Then, by fitting the experimental data of the low-lying mesons, the model parameters
are determined. The involved low-lying mesons in the fitting include $\pi$, $\rho$, $\omega$, $\eta$, $\eta_c(1S)$, $\eta_c(2S)$, $J/\psi(1S)$, $J/\psi(2S)$, $\chi_{c_J}(1P)(J=0,1,2)$, $h_c(1P)$,
totally twelve mesons. With the obtained new set of model parameters, we calculated the high-lying excited-state energy spectrum of charmonium $\chi_{c_J}(2P)(J=0,1,2)$
and $1D$ $c\bar{c}$ mesons. For some exotic states, for example $X(3872)$ can be described well in the unquenched quark model.

The paper is organized as follows. In Sec. \ref{GEM and chiral quark model} the chiral quark model and the GEM are presented. In Sec. \ref{sec3P0}, we introduce the modified
transition operator. The discussion of the results is given in Sec. \ref{Numerical Results}. The last section is devoted to the summary of the present work.

\section{Chiral quark model}
\label{GEM and chiral quark model}
In the nonrelativistic quark model, we obtained the meson spectrum by solving
the Schr\"{o}dinger equation:
\begin{equation}
\label{Hamiltonian1} H \Psi_{M_I M_J}^{IJ} (1,2) =E^{IJ} \Psi_{M_I
M_J}^{IJ} (1,2)\,,
\end{equation}
where $1$, $2$ represents the quark and antiquark. $\Psi_{M_I M_J}^{IJ}(1,2)$ is the wave function of a meson composed of a quark and a antiquark with quantum numbers $IJ^{P}$ and reads,
\begin{align}
\nonumber
& \Psi_{M_I M_J}^{IJ}(1,2) \\
& =\sum_{\alpha}C_{\alpha} \left[
\psi_{l}(\mathbf{r})\chi_{s}(1,2)\right]^{J}_{M_J}
\omega^c(1,2)\phi^I_{M_I}(1,2), \label{PsiIJM}
\end{align}
where $\psi_{l}(\mathbf{r})$, $\chi_{s}(1,2)$,
$\omega^c(1,2)$, $\phi^I(1,2)$ are orbit, spin, color and flavor wave functions, respectively. $\alpha$ denotes the intermediate quantum numbers, $l,s$ and
possible flavor indices. In our calculations, the orbital wave functions is expanded using a set of Gaussians,
\begin{subequations}
\label{radialpart}
\begin{align}
\psi_{lm}(\mathbf{r}) & = \sum_{n=1}^{n_{\rm max}} c_{n}\psi^G_{nlm}(\mathbf{r}),\\
\psi^G_{nlm}(\mathbf{r}) & = N_{nl}r^{l}
e^{-\nu_{n}r^2}Y_{lm}(\hat{\mathbf{r}}),
\end{align}
\end{subequations}
with the Gaussian size parameters chosen according to the
following geometric progression
\begin{equation}\label{gaussiansize}
\nu_{n}=\frac{1}{r^2_n}, \quad r_n=r_1a^{n-1}, \quad
a=\left(\frac{r_{n_{\rm max}}}{r_1}\right)^{\frac{1}{n_{\rm
max}-1}}.
\end{equation}
This procedure enables optimization of the ranges using just a
small number of Gaussians.

At this point, the wave function in Eq.\,\eqref{PsiIJM} is expressed as follows:
\begin{align}
\nonumber
&\Psi_{M_I M_J}^{IJ}(1,2) \\
& =\sum_{n\alpha} C_{\alpha}c_n
 \left[ \psi^G_{nl}(\mathbf{r})\chi_{s}(1,2) \right]^{J}_{M_J}\omega^c(1,2)\phi^I_{M_I}(1,2).\label{Gauss1}
\end{align}
We employ the Rayleigh-Ritz variational principle for solving the Schr\"{o}dinger equation, which leads to a generalized eigenvalue problem due to the non-orthogonality of Gaussians
\begin{subequations}
\label{HEproblem}
\begin{align}
 \sum_{n^{\prime},\alpha^{\prime}} & (H_{n\alpha,n^{\prime}\alpha^{\prime}}^{IJ}
-E^{IJ} N_{n\alpha,n^{\prime}\alpha^{\prime}}^{IJ}) C_{n^{\prime}\alpha^{\prime}}^{IJ} = 0, \\
 &H_{n\alpha,n^{\prime}\alpha^{\prime}}^{IJ} =
  \langle\Phi^{IJ}_{M_I M_J,n\alpha}| H | \Phi^{IJ}_{M_I M_J,n^{\prime}\alpha^{\prime}}\rangle ,\\
 &N_{n\alpha,n^{\prime}\alpha^{\prime}}^{IJ}=
  \langle\Phi^{IJ}_{M_I M_J,n\alpha}|1| \Phi^{IJ}_{M_I M_J,n^{\prime}\alpha^{\prime}}\rangle,
\end{align}
\end{subequations}
with
$\Phi^{IJ}_{M_I M_J,n\alpha} =
[\psi^G_{nl}(\mathbf{r})\chi_{s}(1,2) ]^{J}_{M_J}
\omega^c(1,2)\phi^I_{M_I}(1,2)$,
$C_{n\alpha}^{IJ} = C_{\alpha}c_n$.

We obtain the mass of the four-quark system also by solving the Schr\"{o}dinger equation:
\begin{equation}
    H \, \Psi^{IJ}_{M_IM_J}(4q)=E^{IJ} \Psi^{IJ}_{M_IM_J}(4q),
\end{equation}
where $\Psi^{IJ}_{M_IM_J}(4q)$ is the wave function of the four-quark system, which can be constructed as follows. In our calculations, we only consider the color singlet-singlet
meson-meson picture for the four quark system. First, we write down the wave functions of two meson clusters,
\begin{subequations}
\label{Mesonfunctions}
\begin{align}
\nonumber
&    \Psi^{I_1J_1}_{M_{I_1}M_{J_1}}(1,2)=\sum_{\alpha_1 n_1} {\mathpzc C}^{\alpha_1}_{n_1} \\
    & \times  \left[ \psi^G_{n_1 l_1}(\mathbf{r}_{12})\chi_{s_1}(1,2)\right]^{J_1}_{M_{J_1}}
 \omega^{c_1}(1,2)\phi^{I_1}_{M_{I_1}}(1,2),   \\
&    \Psi^{I_2J_2}_{M_{I_2}M_{J_2}}(3,4)=\sum_{\alpha_2 n_2} {\mathpzc C}^{\alpha_2}_{n_2} \nonumber \\
    & \times \left[ \psi^G_{n_2 l_2}(\mathbf{r}_{34})\chi_{s_2}(3,4)\right]^{J_2}_{M_{J_2}}
    \omega^{c_2}(3,4)\phi^{I_2}_{M_{I_2}}(3,4),
\end{align}
\end{subequations}
then the total wave function of the four-quark state is:
\begin{align}
\label{4qfunctions}
& \Psi^{IJ}_{M_IM_J}(4q)  =  {\cal A} \sum_{L_r}\left[
\Psi^{I_1J_1}(1,2)\Psi^{I_2J_2}(3,4)
     \psi_{L_r}(\mathbf{r}_{1234})\right]^{IJ}_{M_IM_J}    \nonumber \\
\nonumber
  & =  \sum_{\alpha_1\,\alpha_2\,n_1\,n_2\,L_r}
  {\mathpzc C}^{\alpha_1}_{n_1} {\mathpzc C}^{\alpha_2}_{n_2} \bigg[ \left[\psi^G_{n_1 l_1}(\mathbf{r}_{12})\chi_{s_1}(1,2)\right]^{J_1} \nonumber \\
& \quad \times
            \left[\psi^G_{n_2 l_2}(\mathbf{r}_{34})\chi_{s_2}(3,4)\right]^{J_2}
             \psi_{L_r}(\mathbf{r}_{1234})\bigg]^{J}_{M_J} \nonumber \\
 &      \quad  \times \left[\omega^{c_1}(1,2)\omega^{c_2}(3,4)\right]^{[1]}
     \left[\phi^{I_1}(1,2)\phi^{I_2}(3,4)\right]^{I}_{M_I},
\end{align}
Here, ${\cal A}$ is the antisymmetrization operator, if all quarks
(antiquarks) are taken as identical particles, then
\begin{equation}
{\cal A}=\frac{1}{2}(1-P_{13}-P_{24}+P_{13}P_{24}).
\end{equation}
$\psi_{L_r}(\mathbf{r}_{1234})$ is the relative wave function between two clusters, which is also expanded in a set of Gaussians. $L_r$ is the relative orbital angular momentum.

The Hamiltonian of the chiral quark model for the four-quark system consists of three parts: quark rest mass, kinetic energy, potential energy (four-quark system is taken as an example):
\begin{align}
 H & = \sum_{i=1}^4 m_i  +\frac{p_{12}^2}{2\mu_{12}}+\frac{p_{34}^2}{2\mu_{34}}
  +\frac{p_{r}^2}{2\mu_{r}}  \quad  \nonumber \\
  & + \sum_{i<j=1}^4 \left( V_{\rm CON}^{C}(\boldsymbol{r}_{ij})+ V_{\rm OGE}^{C}(\boldsymbol{r}_{ij}) \right. \quad  \nonumber \\
  & \left. + V_{\rm CON}^{SO}(\boldsymbol{r}_{ij}) + V_{\rm OGE}^{SO}(\boldsymbol{r}_{ij}) +\sum_{\chi=\pi,K,\eta} V_{ij}^{\chi}
   +V_{ij}^{\sigma}\right).
\end{align}
Where $m_i$ is the constituent mass of $i$th quark (antiquark). $\frac{\bf{p^2_{ij}}}{2\mu_{ij}}~ (ij=12; 34)$ and $\frac{\bf{p^2_{r}}}{2\mu_{r}}$ represents the inner kinetic of two clusters
and the relative motion kinetic between two clusters, respectively, with
\begin{subequations}
\begin{align}
\bf{p}_{12}&=\frac{m_2\mathbf{p}_1-m_1\mathbf{p}_2}{m_1+m_2}, \\
\mathbf{p}_{34}&=\frac{m_4\mathbf{p}_3-m_3\mathbf{p}_4}{m_3+m_4},  \\
\mathbf{p}_{r}&= \frac{(m_3+m_4)\mathbf{p}_{12}-(m_1+m_2)\mathbf{p}_{34}}{m_1+m_2+m_3+m_4}, \\
\mu_{ij}&=\frac{m_im_j}{m_i+m_j}, \\
\mu_{r}&=\frac{(m_1+m_2)(m_3+m_4)}{m_1+m_2+m_3+m_4}.
\end{align}
\end{subequations}
$V_{\rm CON}^{C}$ and $V_{\rm OGE}^{C}$ are the central parts of the confinement and one-gluon-exchange. $V_{\rm CON}^{SO}$ and $V_{\rm OGE}^{SO}$ are the spin-orbit interaction potential energy.
In our calculations, a quadratic confining potential is adopted. For the mesons, the distance between $q$ and $\bar{q}$ is relatively small, so the difference between the linear potential
and the quadratic potential is very small by adjusting the confinement strengths. Both of them can conform to the linear Regge trajectories for $q\bar{q}$ mesons.
$V_{ij}^{\chi=\pi, K, \eta}$, and $\sigma$ exchange represents the one Goldstone boson exchange. Chiral symmetry suggests dividing quarks into two different sectors:
light quarks ($u$, $d$ and $s$) where the chiral symmetry is spontaneously broken and heavy quarks ($c$ and $b$) where the symmetry is explicitly broken.
The origin of the constituent quark mass can be traced back to the spontaneous breaking of chiral symmetry and consequently constituent quarks should interact through
the exchange of Goldstone bosons. The detailed derivation process can be found in several theoretical papers \cite{Moszkowski,Machleidt}. Here we only show the expressions of these potentials
to save space.

The detailed expressions of the potentials are \cite{Valcarce:2005em}: {\allowdisplaybreaks
\begin{subequations}
\begin{align}
V_{\rm CON}^{C}(\boldsymbol{r}_{ij})&= ( -a_c r_{ij}^2-\Delta ) \boldsymbol{\lambda}_i^c \cdot \boldsymbol{\lambda}_j^c ,  \\
V_{\rm CON}^{\rm SO}(\boldsymbol{r}_{ij})&=\boldsymbol{\lambda}_i^c \cdot \boldsymbol{\lambda}_j^c \cdot \frac{-a_c}{2m_i^2m_j^2}\bigg\{ \bigg((m_i^2+m_j^2)(1-2a_s) \nonumber \\
&+4m_im_j(1-a_s)\bigg)(\boldsymbol{S_{+}} \cdot \boldsymbol{L})+(m_j^2-m_i^2) \nonumber \\
&(1-2a_s)(\boldsymbol{S_{-}} \cdot \boldsymbol{L})\bigg\}, \\
 V_{\rm OGE}^{C}(\boldsymbol{r}_{ij})&= \frac{\alpha_s}{4} \boldsymbol{\lambda}_i^c \cdot \boldsymbol{\lambda}_{j}^c
\left[\frac{1}{r_{ij}}-\frac{2\pi}{3m_im_j}\boldsymbol{\sigma}_i\cdot
\boldsymbol{\sigma}_j
  \delta(\boldsymbol{r}_{ij})\right],  \\
V_{\rm OGE}^{\rm SO}(\boldsymbol{r}_{ij})&=-\frac{1}{16}\cdot\frac{\alpha_s}{m_i^2m_j^2}\boldsymbol{\lambda}_i^c
\cdot \boldsymbol{\lambda}_j^c\big\{\frac{1}{r_{ij}^3}-\frac{e^{-r_{ij}/r_g(\mu)}}{r_{ij}^3}\cdot \nonumber \\
&(1+\frac{r_{ij}}{r_g(\mu)})\big\}\times \bigg\{\bigg((m_i+m_j)^2+2m_im_j\bigg)\nonumber \\
&(\boldsymbol{S_{+}} \cdot \boldsymbol{L})+(m_j^2-m_i^2)(\boldsymbol{S_{-}} \cdot \boldsymbol{L})\bigg\},  \\
\delta{(\boldsymbol{r}_{ij})} & =  \frac{e^{-r_{ij}/r_0(\mu_{ij})}}{4\pi r_{ij}r_0^2(\mu_{ij})}, \mathbf{S}_{\pm}=\mathbf{S}_1\pm \mathbf{S}_2,\\
V_{\pi}(\boldsymbol{r}_{ij})&= \frac{g_{ch}^2}{4\pi}\frac{m_{\pi}^2}{12m_im_j}
  \frac{\Lambda_{\pi}^2}{\Lambda_{\pi}^2-m_{\pi}^2}m_\pi v_{ij}^{\pi}
  \sum_{a=1}^3 \lambda_i^a \lambda_j^a,  \\
V_{K}(\boldsymbol{r}_{ij})&= \frac{g_{ch}^2}{4\pi}\frac{m_{K}^2}{12m_im_j}
  \frac{\Lambda_K^2}{\Lambda_K^2-m_{K}^2}m_K v_{ij}^{K}
  \sum_{a=4}^7 \lambda_i^a \lambda_j^a,   \\
\nonumber V_{\eta} (\boldsymbol{r}_{ij})& =
\frac{g_{ch}^2}{4\pi}\frac{m_{\eta}^2}{12m_im_j}
\frac{\Lambda_{\eta}^2}{\Lambda_{\eta}^2-m_{\eta}^2}m_{\eta}
v_{ij}^{\eta}  \\
 & \quad \times \left[\lambda_i^8 \lambda_j^8 \cos\theta_P
 - \lambda_i^0 \lambda_j^0 \sin \theta_P \right],   \\
v_{ij}^{\chi}(\boldsymbol{r}_{ij}) & =  \left[ Y(m_\chi r_{ij})-
\frac{\Lambda_{\chi}^3}{m_{\chi}^3}Y(\Lambda_{\chi} r_{ij})
\right]
\boldsymbol{\sigma}_i \cdot\boldsymbol{\sigma}_j,\\
V_{\sigma}(\boldsymbol{r}_{ij})&= -\frac{g_{ch}^2}{4\pi}
\frac{\Lambda_{\sigma}^2}{\Lambda_{\sigma}^2-m_{\sigma}^2}m_\sigma \nonumber \\
& \quad \times \left[
 Y(m_\sigma r_{ij})-\frac{\Lambda_{\sigma}}{m_\sigma}Y(\Lambda_{\sigma} r_{ij})\right]  ,
\end{align}
\end{subequations}}
\hspace*{-0.5\parindent}%
where $\mathbf{S}_1$ and $\mathbf{S}_2$ is the spin of the two meson clusters. $Y(x)  =   e^{-x}/x$; $r_0(\mu_{ij}) =s_0/\mu_{ij}$; $\boldsymbol{\sigma}$ are the $SU(2)$
Pauli matrices; $\boldsymbol{\lambda}$, $\boldsymbol{\lambda}^c$ are $SU(3)$ flavor, color Gell-Mann matrices, respectively. The form factor parameter $\Lambda_{\chi}$
$(\chi=\pi, K, \eta, \sigma)$ is introduced to remove the short-range contribution of Goldstone bosons exchanges. $g^2_{ch}/4\pi$ is the chiral coupling constant, determined from
the $\pi$-nucleon coupling; and $\alpha_s$ is an effective scale-dependent running coupling
\cite{Valcarce:2005em},
\begin{equation} \label{alphas}
\alpha_s(\mu_{ij})=\frac{\alpha_0}{\ln\left[(\mu_{ij}^2+\mu_0^2)/\Lambda_0^2\right]}.
\end{equation}
In our calculations, for the two-quark system, besides the central potential energy, the noncentral potential energy is also included. But in the four-quark system calculations,
we find that the influence of the noncentral potential energy on the mass shift of the state is tiny, so it is omitted.

Lastly, we show the model parameters in Table \ref{modelparameters}. In the table, the $\theta_p(^\circ)$ equals $-15$. The angle $\theta_p$ is the mixing angle between $\eta_1$ and $\eta_8$. $|\eta_1\rangle={\rm cos}(\theta_p)|\eta_1\rangle+{\rm sin}(\theta_p)|\eta_8\rangle$, $|\eta^\prime\rangle={\rm sin}(\theta_p)|\eta_1\rangle+{\rm cos}(\theta_p)|\eta_8\rangle$, with $|\eta_1\rangle=(u\bar{u}+d\bar{d}+s\bar{s})/\sqrt{3}$ and $|\eta_8\rangle=(u\bar{u}+d\bar{d}-2s\bar{s})/\sqrt{6}$. What's more, $\Lambda_0$ parameter is  an adjustable parameter to parameterize the running coupling constant, and it has nothing to do with the $\Lambda_{\rm QCD}$. As stated in Ref. \cite{Vijandemodel}, the usual one-loop expression of the running coupling constant will diverges when Q $\rightarrow$ $\Lambda_{\rm QCD}$, so the effective formula of the scale-dependent strong coupling constant is used in chiral quark model.
Need to be noted that, in Ref. \cite{Vijandemodel}, the confinement item takes the screened form $V^{C}_{ij}=\big(-a_c(1-e^{-\mu_c r_{ij}}\big)+\Delta)(\boldsymbol{\lambda}_i^c \cdot \boldsymbol{\lambda}_j^c)$, and in our present calculations, the usual quadratic confinement
$V^{C}_{ij}=( -a_c r_{ij}^2-\Delta ) \boldsymbol{\lambda}_i^c \cdot \boldsymbol{\lambda}_j^c$ is employed, so some parameters are different, such as quark mass, $a_c$ and $\Delta$.
In the nonrelativistic valence quark model, using the model parameters, we calculated the masses of some mesons from light to heavy, the results are shown in the third column of
Table \ref{newspectrum}. It can be seen that most of the ground-state mesons are consistent with the experiment values, but for some excited charmonium states, the quark model
cannot describe them very well.

\begin{table}[!t]
\begin{center}
\caption{ \label{modelparameters} Model parameters, determined by
fitting the meson spectrum, leaving room for unquenching
contributions in the case of light-quark systems.}
\begin{tabular}{llr}
\hline\noalign{\smallskip}
Quark masses   &$m_u=m_d$     &313  \\
   (MeV)       &$m_s$         &536  \\
               &$m_c$         &1728 \\
               &$m_b$         &5112 \\
\hline
Goldstone bosons                &$m_{\pi}$        &0.70  \\
(fm$^{-1} \sim 200\,$MeV )      &$m_{\sigma}$     &3.42  \\
                                &$m_{\eta}$       &2.77  \\
                                &$m_{K}$          &2.51  \\
                                &$\Lambda_{\pi}=\Lambda_{\sigma}$     &4.2  \\
                                &$\Lambda_{\eta}=\Lambda_{K}$         &5.2  \\
                   \cline{2-3}
                               &$g_{ch}^2/(4\pi)$                     &0.54  \\
                               &$\theta_p(^\circ)$                    &-15 \\
\hline
Confinement                    &$a_c$ (MeV fm$^{-2}$)                 &101 \\
                               &$\Delta$ (MeV)                        &-78.3 \\
\hline
OGE                            & $\alpha_0$                           &3.67 \\
                               &$\Lambda_0({\rm fm}^{-1})$            &0.033 \\
                               &$\mu_0$(MeV)                          &36.98 \\
                               &$s_0$(MeV)                            &28.17 \\
\hline
\end{tabular}
\end{center}
\end{table}


\section{The transition operator}
\label{sec3P0}
The $^3P_0$ quark-pair creation model \cite{npB10,
LeYaouanc:1972vsx, LeYaouanc:1973ldf} has been widely applied to
OZI rule allowed two-body strong decays of hadrons
\cite{Capstick:1986bm, Roberts:1992js, Capstick:1993kb,
Page:1995rh, Ackleh:1996yt, Segovia:2012cd}. If the quark and antiquark in the original meson are labeled by 1, 2, and the quark and antiquark ($u\bar{u}$, $d\bar{d}$,
$s\bar{s}$) generated in the vacuum are numbered as 3, 4, the transition operator of the $^3P_0$ model reads:
\begin{align} \label{T0}
T_0 & =-3\, \gamma \sum_m\langle 1m1(-m)|00\rangle\int
d\mathbf{p}_3d\mathbf{p}_4\delta^3(\mathbf{p}_3+\mathbf{p}_4)\nonumber\\
& \quad \times{\cal{Y}}^m_1(\frac{\mathbf{p}_3-\mathbf{p}_4}{2})
\chi^{34}_{1-m}\phi^{34}_0\omega^{34}_0b^\dagger_2(\mathbf{p}_3)d^\dagger_3(\mathbf{p}_4),
\end{align}
where, $\chi^{34}_{1-m},\phi^{34}_0,\omega^{34}_0$ are spin, flavor and color wave functions of the created quark pair, respectively. ${\cal{Y}}^{m}_{1}(\frac{\mathbf{p}_3-\mathbf{p}_4}{2})$ = $pY^m_1(\hat{\mathbf{p}})$ is the solid spherical harmonics. $\gamma$ describes the probability for creating
a quark-antiquark pair with momenta $\mathbf{p}_3$ and $\mathbf{p}_4$ from the vacuum. It is normally determined by fitting the strong decay widths of hadrons.
This yields $\gamma=6.95$ for $u\bar{u}$ and $d\bar{d}$ pair creation, and $\gamma=6.95/\sqrt{3}$ for $s\bar{s}$ pair creation \cite{LeYaouanc:1977gm}.

To reduce the mass shift due to the coupled-channel effects, the transition operator in Eq. (\ref{T0}) should be modified. In Ref. \cite{Chen:2017mug}, two suppression factors are introduced, namely energy damping factor and distance damping factor.
The first factor is $\exp[-r^2/(4f^2)]$ ($\exp[-f^2 p^2]$ in momentum space), here $\mathbf{r}=\mathbf{r_3}-\mathbf{r_4}$ is the distance between the quark and antiquark created in the vacuum, considering the effect of quark-antiquark energy created in the vacuum and it suppresses the contribution from meson-meson states with high energy. What's more, when the distance between the bare meson and a pair of charmed mesons becomes smaller. The energy of tetraquark will increase, and the momentum of the created quark (antiquark) will be large. At this point, the energy damping factor $\exp[-f^2 p^2]$ comes into play, thus the mass shift of the charmed mesons is still suppressed and the convergence is guaranteed.
The second factor is $\exp[-R_{AV}^2/R_0^2]$, which takes into account the effect that the created
quark-antiquark pair should not be far away from the source meson. Here, $R_{AV}$ represents the distance between the created quark-antiquark pair and the source meson. It reads,
\begin{subequations}
\begin{align}
\mathbf{R_{AV}}&= \mathbf{R_A}-\mathbf{R_V};\\
\mathbf{R_A}&=\frac{m_1\mathbf{r_1}+m_2\mathbf{r_2}}{m_1+m_2}; \\
\mathbf{R_V}&= \frac{m_3\mathbf{r_3}+m_4\mathbf{r_4}}{m_3+m_4}=\frac{\mathbf{r_3}+\mathbf{r_4}}{2}~~ (m_3=m_4).
\end{align}
\end{subequations}
So the modified transition operator takes
\begin{align} \label{T1}
T_1&= -3\gamma\sum_{m}\langle 1m1(-m)|00\rangle\int
d\mathbf{r_3}d\mathbf{r_4}(\frac{1}{2\pi})^{\frac{3}{2}}ir2^{-\frac{5}{2}}f^{-5}
\nonumber \\
 & Y_{1m}(\hat{\mathbf{r}})
 {\rm e}^{-\frac{\mathbf{r}^2}{4f^2}}
 {\rm e}^{-\frac{R_{AV}^2}{R_0^2}}\chi_{1-m}^{34}\phi_{0}^{34}
 \omega_{0}^{34}b_3^{\dagger}(\mathbf{r_3})d_4^{\dagger}(\mathbf{r_4}),
\end{align}
By fitting the decay width of $\rho \rightarrow \pi\pi$, and with the requirement that the mass shift is around the $10\%$ of the bare mass, the parameters $f$, $R_0$ and $\gamma$ were fixed,
\begin{equation}\label{para}
\gamma = 32.2,\; \quad f=0.5\,\mbox{fm},\; \quad R_0=1\,\mbox{fm}.
\end{equation}


\section{Numerical Results}
\label{Numerical Results}
In UQM, we obtain the eigenvalues of systems (quark-antiquark plus four-quark components) by solving the Schr\"{o}dinger equation,
\begin{eqnarray}
H\Psi=E\Psi ,
\end{eqnarray}
where $\Psi$ and $H$ is the wave function and the Hamiltonian of the system, respectively. It reads,
\begin{eqnarray}
\Psi=c_1\Psi_{2q}+c_2\Psi_{4q} ~,\\
 H=H_{2q}+H_{4q}+T ~.
\end{eqnarray}
The term $H_{2q}$ only acts on the wave function of two-quark system, $\Psi_{2q}$, and the  $H_{4q}$ only acts on the wave function of four-quark system, $\Psi_{4q}$. The transition operator $T$ is responsible for mixing the quark-antiquark and four-quark components.

In this way, we can get the matrix elements of the Hamiltonian,
\begin{align}
\langle\Psi| & H|\Psi\rangle = \langle
c_1\Psi_{2q}+c_2\Psi_{4q}|H|c_1\Psi_{2q}+c_2\Psi_{4q}\rangle
\nonumber \\
&=c_1^2\langle\Psi_{2q}|H_{2q}|\Psi_{2q}\rangle+c_2^2\langle\Psi_{4q}|H_{4q}|\Psi_{4q}\rangle
\nonumber \\
&\quad+c_1c_2^*\langle\Psi_{4q}|T|\Psi_{2q}\rangle+c_1^*c_2\langle\Psi_{2q}|T^{\dagger}|\Psi_{4q}\rangle,
\end{align}
and the block-matrix structure for the Hamiltonian and overlap
takes,
\begin{equation}
(H)=\left[\begin{array}{cc} (H_{2q}) & (H_{24})\\
(H_{42}) & (H_{4q})
\end{array}
\right],
(N)=\left[\begin{array}{cc} (N_{2q}) & (0)\\
(0) & (N_{4q})
\end{array}
\right] \,,
\end{equation}
with
{\allowdisplaybreaks
\begin{subequations}
\begin{align}
 (H_{2q})&=\langle\Psi_{2q}|H_{2q}|\Psi_{2q}\rangle, \\
 (H_{24})&=\langle\Psi_{4q}|T|\Psi_{2q}\rangle, \\
 (H_{4q})&=\langle\Psi_{4q}|H_{4q}|\Psi_{4q}\rangle,\\
(N_{2q})&=\langle\Psi_{2q}|1|\Psi_{2q}\rangle, \\
(N_{4q})&=\langle\Psi_{4q}|1|\Psi_{4q}\rangle.
\end{align}
\end{subequations}
}
Where $(H_{2q})$ and $(H_{4q})$ is the matrix for the pure two-quark system and pure four-quark system. $(N_{2q})$ and $(N_{4q})$ is their respective overlap matrix. $(H_{24})$ is the coupling matrix of two-quark system and four-quark system.

Finally the eigenvalues ($E_n$) and eigenvectors ($C_n$) of the system are obtained by solving the generalized eigen-problem,
\begin{eqnarray}
\Big[
\begin{array}{c}
(H)-E_n(N)
\end{array}
\Big] \Big[
\begin{array}{c} C_n
\end{array}
\Big]=0. \label{geig}
\end{eqnarray}

By employing the model parameters in Table \ref{modelparameters} and taking the original transition operator $T_0$ in Eq. (\ref{T0}), we calculated the mass shifts for the light
ground-state mesons ($\pi,\rho, \omega,\eta$) \cite{Chen:2017mug} and some charmonium $c\bar{c}$ states \cite{Chen:2023wfp} in our previous work. The results show that for the light ground-state mesons, the coupled-channel effects generate alarmingly large negative mass shifts, and the average value is about $2000$ MeV. For $c\bar{c}$, the mass shifts in \cite{Chen:2023wfp} vary among states, and the average is around $500$ MeV.
Such large mass shift will challenge the validity of the valence quark model as a good zeroth order approximation in describing the low-lying hadron spectrum.

Therefore, we introduced modifications to the transition operator to develop a more realistic unquenching procedure. By adopting the modified transition operator
in Eq. (\ref{T1}), we also demonstrated our new mass shifts for the light ground-state mesons ($\pi,\rho, \omega,\eta$) \cite{Chen:2017mug} and for some charmonium $c\bar{c}$
states \cite{Chen:2023wfp}. For the light ground-state mesons, the mass shifts have been reduced to be just $10\%-25\%$ of a given meson's bare mass, and for $c\bar{c}$,
the unquenching correction is just $1\%-4\%$ of a given meson's bare mass. So the effects of including energy damping factor and distance damping factor on the mass shifts are
relatively stable.

\begin{table}[!t]
\begin{center}
\caption{ \label{newpara} Adjusted quark model parameters.}
\begin{tabular}{llr}
\hline\noalign{\smallskip}
Quark masses   &$m_u=m_d$     &361  \\
   (MeV)       &$m_s$         &477  \\
               &$m_c$         &1700 \\
               &$m_b$         &5112 \\
\hline
Goldstone bosons                &$m_{\pi}$        &0.70  \\
(fm$^{-1} \sim 200\,$MeV )      &$m_{\sigma}$     &3.42  \\
                                &$m_{\eta}$       &2.77  \\
                                &$m_{K}$          &2.51  \\
                                &$\Lambda_{\pi}=\Lambda_{\sigma}$     &4.2  \\
                                &$\Lambda_{\eta}=\Lambda_{K}$         &5.2  \\
                   \cline{2-3}
                               &$g_{ch}^2/(4\pi)$                     &0.54  \\
                               &$\theta_p(^\circ)$                    &-15 \\
\hline
Confinement                    &$a_c$ (MeV fm$^{-2}$)                 &120 \\
                               &$\Delta$ (MeV)                        &-53 \\
\hline
OGE                            &$\alpha_{uu}$                        &0.72 \\
                               &$\alpha_{us}$                        &0.75 \\
                               &$\alpha_{cc}$                        &0.39 \\
                               &$\alpha_{cu}$                        &0.44 \\
                               &$\alpha_{cs}$                        &0.38 \\
                               &$s_0$(MeV)                            &34 \\
\hline
\end{tabular}
\end{center}
\end{table}

What's more, in our previous work Refs. \cite{Chen:2017mug, Chen:2023wfp}, we made minor adjustments to few parameters, which increased the bare masses of mesons. After considering the effects of coupled channels, the mass shifts resulted in a reduction of the state masses, which were then compared with experimental data. However, in the current paper, we have adjusted almost all the model parameters (except the ones related to Goldstone boson exchange), recalculating the masses of the light quark states and charmonium states in the unquenched quark model and comparing them with experimental results, aiming at doing a realistic calculation of meson spectra.

Then, by fitting the experimental data of the low-lying mesons ($\pi$, $\rho$, $\omega$, $\eta$, $\eta_c(1S)$, $\eta_c(2S)$, $J/\psi(1S)$, $J/\psi(2S)$,
$\chi_{c_J}(1P)~(J=0,1,2)$, $h_c(1P)$, totally twelve mesons), the model parameters are determined and given in Table \ref{newpara}. The parameters related to the confinement, one-gluon-exchange potentials and the masses of quarks are all re-adjusted, but the model parameters related to the Goldstone boson exchange potentials are unchanged.
In the nonrelativistic valence quark model, using the new adjusted parameters in Table \ref{newpara}, we re-calculated the masses of some mesons from light to heavy, and listed
them in the fourth column($M_2$) of Table \ref{newspectrum}. By comparing the values in third($M_1$) and fourth columns($M_2$), we found that with the new quark model parameters, the masses of
mesons become larger and the new theoretical threshold information is provided.

\linespread{1.2}
\begin{table}[!t]
\begin{center}
\renewcommand\tabcolsep{10.0pt} 
\caption{ \label{newspectrum} The mass spectrum in the chiral quark model, in comparison with the experimental data \cite{PDG}. $M_1$ represents the mass spectrum with the model parameters in Table \ref{modelparameters}, and $M_2$ represents the mass spectrum with the new adjusted quark model parameters in Table \ref{newpara} (unit: MeV).}
\begin{tabular}{ccccc}
\hline\hline\noalign{\smallskip}
Name             &$J^{P(C)}$    &$M_1$       &$M_2$    &PDG \cite{PDG}\\
\hline
$\pi$             &$0^-$        &134.9        &182.6    &135.0 \\
$K$               &$0^-$        &489.4        &242.6    &493.7 \\
$\rho$            &$1^{--}$     &772.3        &922.6    &775.3 \\
$K^*$             &$1^-$        &913.6        &980.9    &892.0\\
$\omega$          &$1^{--}$     &701.6        &852.4    &782.7 \\
$\eta$            &$0^{-+}$     &669.2        &738.7    &547.9\\
$\phi(1020)$      &$1^{--}$     &1015.9       &1117.9   &1019.5 \\
$D^0$             &$0^-$        &1861.9       &2065.2   &1864.8 \\
$D^{*0}$          &$1^-$        &1980.6       &2162.5   &2006.9\\
$D_s^+$           &$0^-$        &1950.1       &2147.8   &1968.4 \\
$D_s^{*+}$        &$1^-$        &2079.9       &2231.6   &2112.2\\
$B^-$             &$0^-$        &5280.7       &5462.9   &5279.3 \\
$B^{*}$           &$1^-$        &5319.6       &5501.6   &5324.7\\
$B_s^0$           &$0^-$        &5367.4       &5503.4   &5366.9 \\
$B_s^*$           &$1^-$        &5410.2       &5543.9   &5415.4\\
$\eta_c(1S)$      &$0^{-+}$     &2964.4       &3063.4   &2983.9\\
$\eta_c(2S)$      &$0^{-+}$     &3507.8       &3651.2   &3637.5\\
$J/\psi$          &$1^{--}$     &3096.4       &3187.7   &3096.0\\
$\psi(2S)$        &$1^{--}$     &3605.0       &3744.4   &3686.1\\
$\chi_{c_0}(1P)$  &$0^{++}$     &3362.8       &3471.3   &3414.7\\
$\chi_{c_0}(2P)$  &$0^{++}$     &3814.7       &3966.7   &$\chi_{c_0}(3915)?$\\
$\chi_{c_1}(1P)$  &$1^{++}$     &3393.9       &3509.3   &3510.7\\
$\chi_{c_1}(2P)$  &$1^{++}$     &3851.9       &4011.3   &$\chi_{c_1}(3872)?$\\
$\chi_{c_2}(1P)$  &$2^{++}$     &3435.8       &3559.2   &3556.2\\
$\chi_{c_2}(2P)$  &$2^{++}$     &3901.1       &4068.9   &$\chi_{c_2}(3930)?$\\
$h_c(1P)$         &$1^{+-}$     &3416.1       &3535.2   &3525.4\\
$h_c(2P)$         &$1^{+-}$     &3877.4       &4040.4   &$Z_c(3900)$?\\
\hline\hline
\end{tabular}
\end{center}
\end{table}

With the new quark model parameters, as well as the modified transition operator $T_1$ in Eq. (\ref{T1}), we calculated the mass shifts of the light ground-state mesons and some charmonia, the results are given in Table \ref{lightmeson} and Table \ref{charmonium}. In the tables, for the selection of tetraquark channels, some factors need to be considered, such as parity conservation, conservation of angular momentum, isospin conservation and exchange symmetry. For the exchange symmetry, it only applies to the identical quarks.

\begin{table}[htb]
\begin{center}
\renewcommand\arraystretch{1.0}
\caption{ \label{lightmeson} Mass shifts (unit: MeV) computed for non-strange mesons with quantum numbers $IJ^-(I=0,1;J=0,1)$ with the new quark model parameters and the modified transition operator $T_{1}$ in Eq.\,\eqref{T1}: $f=0.5$\,fm, $\gamma=32.2$, $R_0=1$\,fm. ($\eta$ is an isospin 0 partner to the pion.)}
\begin{tabular}{lccccc} \hline
$(IJ^P)$ & $\pi(10^-)$ & $\rho(11^-)$ &  $\omega(01^-)$ &  $\eta(00^-)$ \\ \hline
bare mass (Theo.) & 182.6   &  922.6      &  852.4   & 738.7 \\ \hline
$\pi\pi$          & ...       & $-32.9$     & ...        &  ...  \\
$\pi\rho$         & $-13.9$ & ...           & $-37.2$  & ...          \\
$\pi\omega$       & ...       & $-15.1$     &  ...       &  ...  \\
$\eta\rho$        & ...       & $-9.7$      &  ...      &  ...  \\
$\rho\rho$        & ...       & $-31.4$     &  ...      &  $-32.8$  \\
$\rho\omega$      & $-11.9$ &  ...          &  ...       &  ... \\
$\eta\omega$      &  ...      &  ...          &  $-9.2$  &  ... \\
$\omega\omega$    &  ...      &  ...          &  ...       & $-11.5$ \\
$K \bar{K}$       &  ...      &  $-6.0$     &  $-3.6$  &  ... \\
$K\bar{K}^{\star}(\bar{K}K^{\star})$ & $-4.1$ & $-8.4$  &  $-7.4$  & $-9.2$ \\
$K^{\star}\bar{K}^{\star}$           & $-7.3$  &  $-21.5$  &  $-19.9$  &  $-14.5$  \\
\hline
Total mass shift & ~$-37.2$~  & ~$-125.0$~  & ~$-77.3$~  & ~$-68.0$~  \\
\hline
 Unquenched mass &145.4 &797.6 &775.1 &670.7 \\
 \hline
 Exp &139 &772 &782 &547 \\
  \hline
\end{tabular}
\end{center}
\end{table}

\begin{table*}[!t]
\begin{center}
\caption{ \label{charmonium} Mass shifts computed for $c\bar{c}$ charmonium mesons with the new quark model parameters and the modified transition operator
$T_{1}$ in Eq.\,\eqref{T1}: $f=0.5$\,fm, $\gamma=32.2$, $R_0=1$\,fm. (Units of MeV)}
\begin{tabular}{cccccccccccccccc}
\hline \hline \multicolumn{3}{c}{Bare $c\bar{c}$ state}
&\multicolumn{10}{c}{Mass
shifts by channels} &\multicolumn{3}{c}{$c\bar{c}+qq\bar{q}\bar{q}$} \\
\cline{1-3}\cline{5-13}\cline{15-16}
 State$(n^{2S+1}L_J)$ &Bare mass &Exp &~~~ &$D\bar{D}$ &$D\bar{D^*}$ &$D^*\bar{D}$ &$D^*\bar{D^*}$ &$D_s\bar{D_s}$ &$D_s\bar{D}_s^*$ &$D^*_s\bar{D}_s$ &$D^*_s\bar{D}_s^*$ &Total & & &Unquenched mass \\
\hline
$\eta_c(1S)(1 ^1S_0)$        &3063.4 &2983.9             & &...    &-11.8  &-11.8  &-22.7  &...   &-2.9  &-2.9   &-5.7  &-57.8    && &3005.6\\
$\eta_c(2S)(2^1S_0)$         &3651.2 &3637.5             & &...    &-22.1  &-22.1  &-40.4  &...   &-4.0  &-4.0   &-7.7  &-100.3   && &3550.9\\
$J/\psi(1S)(1^3S_1)$         &3187.7 &3096.0             & &-4.7   &-9.1  &-9.1  &-30.8  &-1.1  &-2.2  &-2.2   &-7.6  &-66.8    && &3120.9\\
$\psi(2S)(2^3S_1)$           &3744.4 &3686.1             & &-9.1  &-16.5  &-16.5  &-52.4  &-1.5  &-2.9  &-2.9   &-9.8 &-111.6   && &3632.8\\
$\chi_{c_0}(1P)(1^3P_0)(S+D)$ &3471.3 &3414.7            & &-13.6  &...    &...    &-56.8   &-2.7  &...   &...    &-12.5  &-85.6    && &3385.7\\
$\chi_{c_1}(1P)(1^3P_1)(S+D)$ &3509.3 &3510.7             & &...    &-16.5  &-16.5  &-41.9    &...   &-3.4  &-3.4   &-9.2   &-90.9    && &3418.4\\
$\chi_{c_2}(1P)(1^3P_2)(S+D)$ &3559.2 &3556.2             & &-10.0    &-14.1    &-14.1    &-42.7  &-2.1   &-3.0   &-3.0   &-9.0  &-98.0    && &3461.2\\
$h_c(1P)(1^1P_1)(S+D)$    &3535.2 &3525.4              & &...    &-20.0   &-20.0   &-37.3  &...   &-4.1  &-4.1   &-8.2  &-93.7    && &3441.5\\
$\chi_{c_0}(2P)(2^3P_0)(S+D)$ &3966.7 &$\chi_{c_0}(3915)$? & &-29.0  &...    &...    &-78.9   &-3.7  &...   &...    &-13.8  &-125.4    && &3841.3\\
$\chi_{c_1}(2P)(2^3P_1)(S+D)$ &4011.3 &$\chi_{c_1}(3872)$?& &...    &-29.0  &-29.0  &-58.2    &...   &-4.2  &-4.2  &-10.1   &-134.7    && &3876.6\\
$\chi_{c_2}(2P)(2^3P_2)(S+D)$ &4068.9 &$\chi_{c_2}(3930)$?& &-18.2    &-22.4    &-22.4    &-66.4  &-2.5   &-3.5   &-3.5    &-10.5  &-149.4    && &3919.5\\
$h_c(2P)(2^1P_1)(S+D)$    &4040.4 &$Z_c(3930)$?       & &...    &-33.1  &-33.1  &-56.1  &...   &-4.9  &-4.9   &-9.3  &-141.4    && &3899.0\\
$\eta_{c_2}(1D)(1^1D_2)$     &3824.9 &?               & &...    &-7.9   &-7.9   &-13.6  &...   &-1.3  &-1.3   &-2.5  &-34.5    && &3790.4\\
$\psi(1D)(1^3D_1)$   &3799.8 &$\psi(3770)$?      & &-14.5  &-6.2   &-6.2   &-4.3   &-2.3  &-1.1  &-1.1   &-0.8  &-36.5    && &3763.3\\
$\psi_2(1D)(1^3D_2)$         &3817.1 &$\psi_2(3823)$?    & &...    &-11.6  &-11.6  &-6.9   &...   &-2.0  &-2.0   &-1.2  &-35.3    && &3781.8\\
$\psi_3(1D)(1^3D_3)$         &3839.8 &$\psi_3(3842)$?    & &...    &...    &...    &-28.4  &...   &...   &...    &-5.0  &-33.4    && &3806.4\\
 \hline \hline
\end{tabular}
\end{center}
\end{table*}

From Table \ref{lightmeson}, we can find that with the new set of the quark model parameters, the bare masses of the light ground-state mesons are increased but the mass shifts
are not very sensitive to these model parameters, compared with our previous work \cite{Chen:2017mug}. Eventually, the unquenched masses of mesons are well consistent with the
experimental values.

There exist one open channel in our calculations. The mass of $\rho$ meson is larger than the sum of masses of two pions and it can decay to $\pi\pi$.
For the open channel, which means that final state energy is lower than the bare mass of the meson, the mass shift of the state will change with the Gaussian distribution of the relative motion between two mesons. Especially,
we find that the mass shift will change with the increasing of spatial volume periodically. In our calculations, we picked up the biggest mass shift as
the contributions of this open channel. For $\pi\pi$ state, it is a scattering one, and it has the discrete energy levels which will change with the varying
Gaussian distribution in the theoretical calculations because the limitation of finite volume. When considering the coupling of the $\pi\pi$ and $\rho$,
the strength of coupling will be increased as one of the energy of $\pi\pi$ state is close to that of $\rho$, and the induced mass shift will become bigger.
We take the biggest one as the mass shift of the state $\rho$ to the $\pi\pi$ state. Besides, if we expand the space further with higher $r_n$ values,
the same biggest mass shift will be repeated. From the table, we can also find that for the open channel $\pi\pi$, the mass shift is larger than the other close
channels. From the Table \ref{newspectrum}, we find that $M_2$ of the $c\bar{c}$ states are all below 4100 MeV, and the sums of the masses of $D$ and $D^{(*)}$ are
larger than 4100 MeV. So for the charmonium mesons, there exits no open channels with the new adjusted quark model parameters. In the case of these processes (close channel), the mass shifts will not vary with an increase in space.

The results for the mass spectrum of the $c\bar{c}$ charmonium mesons with the new quark model parameters are shown in Table \ref{charmonium}.  For the ground-state
$\eta_c(1S)$ and $J/\psi(1S)$, the experimental masses are fitted well. For the excited $\eta_c(2S)$ and $J/\psi(2S)$, the theoretical unquenched masses are also close
to the experiment values. The fitted masses for $\chi_{c_J}(1P)(J=0,1,2)$ and $h_c(1P)$ are a little bigger, about 90 MeV. The situation gets better for
$\chi_{c_J}(2P)~(J=0,1,2)$ and $h_c(2P)$. Especially for $\chi_{c_1}(2P)$, the unquenched mass is 3876.6 MeV, which agrees with experimental value of the exotic state
$X(3872)$ very well. Besides, in our work, $\chi_{c_2}(2P)$ has a theoretical mass of 3919.5 MeV, and the mass is very close to the exotic state $\chi_{c_2}(3930)$.
For higher charmonium $1D$ states, for example, the mass of $1^1D_2$ is about 3790.4 MeV. In the future, we look forward to having more experimental data about this state.
$\psi(3770)$ is also described very well in the unquenched quark model, and it may be a good candidate of $1^3D_1$ state, with the unquenched mass 3763.3 MeV.
What's more, the charmonium states $\psi_2(3823)$ and $\psi_3(3842)$ are very likely candidates of $1^3D_2$ and $1^3D_3$, with the theoretical unquenched masses 3781.8
and 3806.4 MeV, respectively.

Need to be noted that, in our calculations, the angular momentum for the two mesons $l_1$ and $l_2$ in Eq. (\ref{Mesonfunctions}) equals zero. So for the charmonium
$1S$, $2S$ and $1D$ states, the relative angular momentum $L_r$ between two mesons in Eq. (\ref{4qfunctions}) may equal 1 ($P$ wave) or 3 ($F$ wave) by considering the parity conservation. Here we only
consider $L_r$ equals 1 for a simplification. For the $1P$ and $2P$ states, we not only consider the $L_r=0$, but also the $L_r=2$. In Table \ref{SDwave},
We demonstrated the mass shifts of $\chi_{c_1}(2P)$ state when the relative angular momentum $L_r$ is in $S$-wave and $D$-wave, respectively. From the table,
one can see that the mass shifts from $D$-wave $D\bar{D}^*$ and  $D_s\bar{D}_s^*$ are smaller than that from corresponding $S$-wave states. For the contributions
from $D$-wave $D^*\bar{D}^*$ and $D_s^*\bar{D}_s^*$ are larger than the corresponding $S$-wave states. The other $1P$ and $2P$ states also follow the similar patterns.
Generally the $D$-wave channels have larger energies than those of $S$-wave channels, and they should have less contribution to the mass of the $c\bar{c}$ state.
The inversion of the contribution from $S$- and $D$-wave channels may lead to the problem of convergency. Further research are expected. And our conclusions
are consistent with that of Ref. \cite{Tan:2019qwe}.

\begin{table*}[!t]
\begin{center}
\caption{ \label{SDwave}   Mass shifts computed for $\chi_{c_1}(2P)$ state when the relative angular momentum $L_r$ is in S-wave and D-wave, respectively. (Units of MeV)}
\begin{tabular}{cccccccccccccccc}
\hline \hline \multicolumn{3}{c}{Bare $c\bar{c}$ state}
&\multicolumn{10}{c}{Mass
shifts by channels} &\multicolumn{3}{c}{$c\bar{c}+qq\bar{q}\bar{q}$} \\
\cline{1-3}\cline{5-13}\cline{15-16}
 State$(n^{2S+1}L_J)$ &Bare mass &Exp &~~~ &$D\bar{D}$ &$D\bar{D^*}$ &$D^*\bar{D}$ &$D^*\bar{D^*}$ &$D_s\bar{D_s}$ &$D_s\bar{D}_s^*$ &$D^*_s\bar{D}_s$ &$D^*_s\bar{D}_s^*$ &Total & & &Unquenched mass \\
\hline
$\chi_{c_1}(2P)(2^3P_1)(S)$ &4011.3 &$\chi_{c_1}(3872)$?& &...    &-17.5 &-17.5  &...    &...   &-2.4  &-2.4  &...   &-39.8    && &\\
$\chi_{c_1}(2P)(2^3P_1)(D)$ &4011.3 &$\chi_{c_1}(3872)$?& &...    &-11.5 &-11.5  &-58.2    &...  &-1.8  &-1.8  &-10.1  &-94.9    && &3876.6\\
 \hline \hline
\end{tabular}
\end{center}
\end{table*}

\begin{table}[!t]
\renewcommand\arraystretch{1.0}
\caption{\label{qqfractions} Fractions (\%) of two- and four-quark components for the light ground-state mesons in the unquenched quark model.}
\setlength{\tabcolsep}{5mm}{
\begin{tabular}{l|cccc}\hline
                & $\pi$ & $\rho$ & $\omega$ & $\eta$ \\\hline
Bare $q \bar q$ &98.3  &39.2   &90    &95.8 \\\hline
$\pi\pi$        &...   &48.4   &...   &... \\
$\pi \rho$      &0.7   &...    &4.6   &... \\
$\pi \omega$    &...   &2.8    &...   &... \\
$\eta \rho$     &...   &0.7    &...   &...\\
$\eta \omega$   &...   &...    &0.7   &... \\
$\rho\rho$      &...   &2.1    &...   &1.9 \\
$\rho\omega$    &0.5   &...    &...   &...  \\
$\omega\omega$  &...   &...    &...   &0.7 \\
$KK$            &...   &4.4    &2.7   &...\\
$K K^\ast$      &0.2   &1      &0.8   &0.8 \\
$K^\ast K^\ast$ &0.3   &1.4    &1.2   &0.8   \\\hline
\end{tabular}}
\end{table}

\begin{table*}[!t]
\renewcommand\arraystretch{1.0}
\caption{\label{ccfractions} Fractions (\%) of two- and four-quark components for the charmonium mesons in the unquenched quark model.}
\setlength{\tabcolsep}{3.7mm}{
\begin{tabular}{cccccccccc}
\hline \hline
                             &Bare $q\bar{q}$ &$D\bar{D}$ &$D\bar{D^*}$ &$D^*\bar{D}$ &$D^*\bar{D^*}$ &$D_s\bar{D_s}$ &$D_s\bar{D}_s^*$ &$D^*_s\bar{D}_s$ &$D^*_s\bar{D}_s^*$  \\
\hline
$\eta_c(1S)(1 ^1S_0)$        &97   &...    &0.7    &0.7    &1.1    &...   &0.1   &0.1   &0.3     \\
$\eta_c(2S)(2^1S_0)$         &90.6 &...    &2.3    &2.3    &3.6    &...   &0.3   &0.3   &0.6   \\
$J/\psi(1S)(1^3S_1)$         &95.8 &0.3    &0.6    &0.6    &1.9    &0.1   &0.1   &0.1   &0.5   \\
$\psi(2S)(2^3S_1)$           &87.8 &1.3    &2.0    &2.0    &5.2    &0.2   &0.3   &0.3   &0.9   \\
$\chi_{c_0}(1P)(1^3P_0)(S+D)$&93.9 &1.3    &...    &...    &3.8    &0.2   &...   &...   &0.8      \\
$\chi_{c_1}(1P)(1^3P_1)(S+D)$&93.2 &...    &1.4    &1.4    &3.0    &...   &0.2   &0.2   &0.6     \\
$\chi_{c_2}(1P)(1^3P_2)(S+D)$&91.8 &0.9    &1.2    &1.2    &3.6    &0.2   &0.2   &0.2   &0.7 \\
$h_c(1P)(1^1P_1)(S+D)$       &92.6 &...    &1.7    &1.7    &2.9    &...   &0.3   &0.3   &0.5     \\
$\chi_{c_0}(2P)(2^3P_0)(S+D)$&81   &7.9    &...    &...    &9.1    &0.6   &...   &...   &1.4     \\
$\chi_{c_1}(2P)(2^3P_1)(S+D)$&78.4 &...    &6.2    &6.2    &7.1    &...   &0.5   &0.5   &1.1    \\
$\chi_{c_2}(2P)(2^3P_2)(S+D)$&72.2 &5.2    &4.3    &4.3    &11.3   &0.4   &0.5   &0.5   &1.3     \\
$h_c(2P)(2^1P_1)(S+D)$       &75.3 &...    &6.9    &6.9    &8.4    &...   &0.7   &0.7   &1.1    \\
$\eta_{c_2}(1D)(1^1D_2)$     &95.3 &...    &1.3   &1.3   &1.7  &...   &0.1  &0.1   &0.2  \\
$\psi(1D)(1^3D_1)$           &94.2 &2.8  &0.9   &0.9   &0.5   &0.3  &0.1  &0.1   &0.2    \\
$\psi_2(1D)(1^3D_2)$         &95 &...    &1.8  &1.8  &0.9   &...   &0.2  &0.2   &0.1     \\
$\psi_3(1D)(1^3D_3)$         &95.9 &...    &...    &...    &3.6  &...   &...   &...    &0.5      \\
 \hline \hline
\end{tabular}}
\end{table*}

Further, we analyze the fractions of the two-quark ($q\bar{q}$) system and the four-quark (meson-meson) system for light ground-states in Table \ref{qqfractions} and
for the charmonium states in Table \ref{ccfractions}. From Table \ref{qqfractions}, we can see that the probability fractions of $q\bar{q}$ components are all over $90\%$.
But for $\rho$ meson, the $q\bar{q}$ component accounts for $39.2\%$, because it can decay to the open channel $\pi\pi$. This open channel makes the largest mass shift
contribution and the fraction of four-quark components $\pi\pi$ are rather large, $48.4\%$.
By the way, the picture of rho meson in UQM is that a quark-antiquark core surrounded by meson cloud. The decay constant and electromagnetic form factor mainly depend on the part of the wave function in the core, so we expect that these quantities will not be changed in UQM.

From Table \ref{ccfractions}, we find that the dominant components of charmonium states are $c\bar{c}$, from $70\% \sim 97\%$ in the unquenched quark model.
For $\chi_{c_J}(1P)~(J=0,1,2)$ and $h_c(1P)$ state, the dominant component $c\bar{c}$ accounts for larger than $90\%$. And the main four-quark components are all
$D^*\bar{D^*}$, $\sim 3\%$. For $\chi_{c_1}(2P)$ state, the unquenched mass agrees with the experimental value of $X(3872)$ very well. The picture of the exotic state
$X(3872)$ is $c\bar{c}$ state mixed with four-quark components. The dominant component of $X(3872)$ is still $c\bar{c}$, about $78.4\%$, and the fraction of
$D\bar{D^*}+D^*\bar{D}$ is about $12.4\%$, $D^*\bar{D^*}$, $7.1\%$, $D_s\bar{D}_s^*+D^*_s\bar{D}_s$, $1\%$ and $D^*_s\bar{D}_s^*$, $1.1\%$.
Our results of $X(3872)$ are consistent qualitatively with some previous work \cite{Tan:2019qwe,Achasov:2017kni,Kang:2016jxw,Ferretti:2013faa}. But in other work,
the dominant components of $X(3872)$ are meson-meson ones, and the fraction of $c\bar{c}$ is small, for example, $7\% \sim 32\%$ in Ref. \cite{Ortega:2009hj},
$7.5\% \sim 11.2\%$ in Ref. \cite{Coito:2012vf}, $14.7\%$ in Ref. \cite{Ferretti:2018tco}. And Kalashnikova got a little large fraction of $c\bar{c}$, about
$54.3\%$ \cite{Kalashnikova:2005ui}. Furthermore, in our work, we also conducted calculations on the four-quark system $\bar{D}^*D$ and did not find any bound state although there is an attraction between $\bar{D}^*$ and $D$. Therefore, in our present framework of the unquenched quark, we consider that $X(3872)$ may be classified as a member of the charmonium family $\chi_{c_1}(2P)$.

For $\chi_{c_2}(3930)$, the candidate of $\chi_{c_2}(2P)$ state in the unquenched quark model, the dominant constituent is also $c\bar{c}$, about $72.2\%$.
The largest four-quark component is $D^*\bar{D^*}$, $11.3\%$, and the $D^{(*)}_s\bar{D}_s^{(*)}$ occupied a very small percentage. For higher excited states $1D$,
the fraction of $c\bar{c}$ is much higher than that of $\chi_{c_J}~(2P)(J=0,1,2)$ and $h_c(2P)$ states, to be $95\%$. For $1^1D_2$ and $1^3D_2$, the main four-quark component
is $D\bar{D^*}+D^*\bar{D}$, $2.6\%$ and $3.6\%$, respectively. For $1^3D_1$, the main four-quark component is $D\bar{D}$, $2.8\%$. For $1^3D_3$, $D^*\bar{D^*}$ accounts for
$3.6\%$. 


\section{Summary}
\label{epilogue}
To give a unified description of the ordinary meson and exotic states in experiments, a new quark model $\textendash$ the unquenched quark model is developed.
As a preliminary work, we calculated the unquenched masses of the ordinary light ground-state mesons ($\pi$, $\rho$, $\omega$, $\eta$), as well as the some charmnonium
$c\bar{c}$ states in UQM.

In UQM, the coupling of the two quark component and the high Fock four-quark component is considered. For the four-quark component, there are different configurations,
such as meson-meson structure, hidden-color structure, diquark-antidiquark structure, and so on. In our present work, only four-quark components in meson-meson structure
are taken into account. Of course, in the future work, the effects and the convergence of the more configuration of the four-quark components will be investigated.

The modified transition operator which relates the valence part to the high Fock components is also applied. Two simple, physically motivated improvements are introduced.
One is to suppress the contribution from the intermediate dressing-states with large momentum. Another is to favor the quark-antiquark creation near the source hadron.
With these improvements, the alarmingly mass shifts are reduced and the success of valence quark model in describing low-lying spectrum of meson is kept.

Further, by fitting the experimental values of $\pi$, $\rho$, $\omega$, $\eta$, $\eta_c(1S)$, $\eta_c(2S)$, $J/\psi(1S)$, $J/\psi(2S)$, $\chi_{c_J}(1P)~(J=0,1,2)$, $h_c(1P)$,
totally twelve mesons, we obtained a set new quark model parameters. We calculated the high excited $c\bar{c}$ states with the new model parameters in UQM, aiming to explain
some exotic states observed in experiments. Here, some well-known exotic states can be described very well. At the same time, the masses of light ground-state mesons
and low-lying charmnia, $\eta_c$ and $J/\psi$ are reproduced well. For example, our calculation shows that the unquenched mass of $\chi_{c_1}(2P)$ is very close to
the experimental value of $X(3872)$, and the dominant component of $X(3872)$ is $c\bar{c}$, about $78.4\%$. $\chi_{c_2}(2P)$ is a good candidate of $\chi_{c_2}(3930)$,
the dominant constituent is also $c\bar{c}$, about $72.2\%$. $\psi(3770)$ is very likely to be the charmonium $1^3D_1$ state. $\psi_2(3823)$ and $\psi_3(3842)$ may be
the candidate of $1^3D_2$ and $1^3D_3$ states, respectively. All of the $1D$ states are the $c\bar{c}$ dominant states, with the fractions about $95\%$.

So the unquenched quark model is a promising phenomenological method to unify the description of ordinary
mesons and exotic mesons. However there are still some problems with the convergency of the Fock expansion. Further improvement of the transition operator are expected.
With the accumulation of experiment data, it will further help us verify the reasonability of the improvements of the quark model.

\acknowledgments
This work is partly supported by the National Natural Science Foundation of China under Grants No. 12205125, No. 11847145, No. 12205249 and No. 11865019, and also supported
by the Natural Science Foundation of Jiangsu Province under Grants No. BK20221166.


\end{document}